\documentclass[pra,aps,showpacs,superscriptaddress,twocolumn]{revtex4-1}
\usepackage{amsmath}
\usepackage{amsfonts}
\usepackage{amssymb}
\usepackage{mathrsfs}
\usepackage{epsfig}
\usepackage{bm}
\usepackage{amsbsy}
\usepackage{color}
\usepackage{epstopdf}

\begin{document}

\newcommand{\ket}[1]{\vert#1\rangle}
\newcommand{\bra}[1]{\langle#1\vert}
\newcommand{\braket}[2]{\langle#1\vert#2\rangle}
\newcommand{\ketbra}[2]{\vert#1\rangle\langle#2\vert}
\newcommand{\braketM}[3]{\langle#1\vert#2\vert#3\rangle}
\newcommand{\modq}[1]{\vert#1\vert^2}
\newcommand{\modulo}[1]{\vert#1\vert}
\newcommand{\mean}[2]{\langle#1\rangle_{#2}}
\newcommand{\vet}[1]{\overrightarrow{#1}}
\newcommand{\mauro}[1]{{\color{blue}{\textbf{#1}}}}
\newcommand{\thomas}[1]{{\color{green}{\textbf{#1}}}}
\newcommand{\nicola}[1]{{\color{blue}{#1}}}
\newcommand{\nicoladue}[1]{{\color{red}{#1}}}

\title{Structural change of vortex patterns in anisotropic
  Bose-Einstein condensates}

\author{N. Lo Gullo} 

\author{Th. Busch} \affiliation{Department of Physics, University
  College Cork, Cork, Republic of Ireland}

\author{M. Paternostro} \affiliation{School of Mathematics and
  Physics, Queen's University, Belfast BT7 1NN, United Kingdom}
\date{\today}

\begin{abstract}
  We study the changes in the spatial distribution of vortices in a
  rotating Bose-Einstein condensate due to an increasing eccentricity
  of the trapping potential. By breaking the rotational symmetry, the
  vortex system undergoes a rich variety of structural changes,
  including the formation of zig-zag and linear configurations. These
  spatial re-arrangements are well signaled by the change in the
  behavior of the vortex-pattern eigenmodes against the eccentricity
  parameter. This behaviour allows to actively control the
    distribution of vorticity in many-body systems and opens the
    possibility to study interactions between quantum vortices over a
    large range of parameters.
\end{abstract}

\maketitle

The superfluid nature of atomic Bose Einstein condensates (BECs) is one of
the most striking manifestations of quantum mechanics on a macroscopic
scale.  Its tell-tale sign, the formation of quantized vortices, has
been extensively studied in recent years and led to significant
progress in understanding the phenomenon \cite{fetterRMP}. However, to
create these topological defects one usually needs to break the
rotational symmetry of the condensate. This requires a significant
external disturbance through, for example, optical phase imprinting
techniques \cite{phaseimp} or stirring laser fields that allow to
excite quadrupole-mode resonances \cite{quadrupole}. The latter one is
similar to the classical {\sl rotating bucket} method that creates a
vortex by rotating a bucket full of water and was first used in
experimental studies of superfluid ${}^4$He~\cite{bucket}. The
validity of such an analogy is certainly limited, given that the non
superfluid component in superfluid ${}^4$He experiences friction due
to its relative motion with respect to the walls.  As in the classical
case, this implies a transfer of energy and angular momentum from the
walls of the bucket to the superfluid part through the non-superfluid
component.  In a trapped BEC the latter is always negligible and the
transfer has to be made through the excitation of normal modes,
usually quadrupole ones.
A BEC thus reacts to a large amount of angular momentum by creating
many vortices with winding number equal to one~\cite{fetterring},
which arrange themselves in geometrically defined spatial patterns. At
large vortex density, these structures mimic the celebrated Abrikosov
lattice~\cite{abri}. For harmonically trapped alkali
  condensates 
this was first observed in the seminal experiment by Abo-Shaeer {\it
  et al.}  \cite{quadrupole}, where more than $100$ vortices formed a
triangular-shaped lattice with a few seconds lifetime.

Recently, numerical evidence has been provided that the vortex pattern
of a 2D BEC in an in-plane anisotropic rotating trap can undergo
structural changes as a function of the eccentricity. Specifically, in
Ref.~\cite{suzanne} it has been shown that, for modest changes in the
eccentricity, an {\it off-line} configuration (typical for an
Abrikosov lattice) can change into a linear one.  While this bears
analogies with the case of ionic crystals~\cite{3ions}, the
characterization of structural changes in anisotropic and rotating
BECs remains largely unexplored. Most of the existing literature
focuses on the limit of large numbers of vortices for either a
symmetric trap~\cite{colossi,JILA}
or very high angular frequencies, which leads to stripe-shaped vortex
patterns~\cite{stripe,asymmetric, stripeII,fetterLLL}.
Although the case of medium vorticity has been addressed, the role of
external forces on the dynamics of the vortex structures still awaits
a systematic
approach~\cite{fetterell,lattice,vortexlattice,campbell}.
Yet, understanding how vortices behave under external perturbations is
a pre-requisite for harnessing the quantum properties of vortex
patterns.
Here we present a significant contribution to advance these aims by
studying the behavior of finite-sized vortex patterns in 2D BECs
confined within a rotating anisotropic trap. In particular we
investigate in detail the effects of the eccentricity on the spatial
distribution of the vortices. By minimizing the eccentricity-dependent
interaction potential between vortices, we show
that the vortex configuration

undergoes structural changes as the eccentricity parameter is
varied. A hydrodynamical approach to the description of the superfluid
motion allows us to identify the eigenmodes of the vortex-patterns and
connect the appearance of discontinuities with the transition points
between different structures. In fact, the modes suggest that the
change in the equilibrium positions of the vortices is due to the
re-arrangement of the superfluid velocity field.


\section{ Vortex pattern} 
We consider the pattern of vortices in the ground state of a BEC held
in a rotating trapping potential. The ground state is found by
minimizing the energy functional~\cite{confenergy}
\begin{equation}
{\cal E}[\Psi,\Psi^{*}]{=}\!\!\int\!\!d^3{\bf
  r}\bigg[\frac{\hbar^2|\nabla\Psi|^2}{2 m}{+}V({\bf
  r})|\Psi|^2{+}\frac{Ng|\Psi|^4}{2}{-}\Psi^*({\bm
  \Omega}{\cdot}{\hat{\bf L}})\Psi\bigg]
  \end{equation} 
where $\Psi$ is the normalized order parameter of the condensate (its
dependence on ${\bf r}$ is omitted for ease of notation), $V({\bf r})$
is the trapping potential, $m$ is the atomic mass, $N$ is the number
of atoms, $g=4\pi\hbar^2{a}/m$ is the inter-atomic interaction energy
volume determined by the s-wave scattering length $a$, ${\bm \Omega}$
is the rotation frequency vector of the condensate and $\hat{\bf L}$
is the angular momentum operator. The function $\Psi$ minimizing
${\cal E}$ has been studied both numerically and analytically under
different working assumptions such as the Thomas-Fermi (TF)
approximation ~\cite{confenergy},
the lowest-Landau-level (LLL) approximation~\cite{lll} or the
  limit of very weak interactions~\cite{lattice}. The first usually
corresponds to the requirement of a very large number of particles, so
that the kinetic
energy associated with $\nabla |\Psi_{NS}|$ (with $\Psi_{NS}$
representing the non-singular part of the order parameter) can be
neglected in favor of the boson-boson interaction. In the LLL
approximation, on the other hand, the main contribution to the energy
stems from the centrifugal term and $\Psi$ is well described
by {means of} single-particle wave-functions.  Finally, in the
  limit of weak interactions the healing length becomes large and even
  under strong rotation only a small number of large vortices
  nucleate~\cite{lattice}.

Here we consider a BEC in a harmonic trap rotating about its $z$-axis,
which is also the direction of tight-confinement, so that $\Psi$
can be factorized into an axial part (the ground state of a
harmonic potential) and an in-plane one, $\psi(x,y)$. We call
$\omega_{j}$ ($j{=}x,y$) the trapping frequency along axis $j$ of the
trap and introduce the eccentricity parameter $\lambda=\omega_y /
\omega_x$.

We are now in a position to minimize ${\cal E}$ in the TF limit.
For a set value of ${0{<}\lambda{\le}1}$, we call
$\Omega_{N_v}(\lambda)$ the minimum angular frequency of the trap
which allows for $N_v$ vortices in the state
which minimizes $E[\Psi,\Psi^{*}]$, while ${\bf r}_i$ is the position
of the $i^{\text{th}}$ vortex in the frame rotating with the
condensate. By introducing $|{\bf
  r}_i|_\lambda^2=x_i^2+\lambda^{2}y_i^2$, the energy of the vortex
pattern can be written as $U{=}U_T+U_I$ with~\cite{francospagnoli}
\begin{equation}
  \label{eq:inter}
\begin{aligned}
  U_T{=}\frac{\pi\rho_0(\lambda)}{(1+\lambda^2)}
         \sum_{i=1}^{N_v}|{\bf r}_{i}|_{\lambda}^2,\,
  U_I{=}{-}\pi\rho_0(\lambda)\sum_{i=1}^{N_v}\!\sum_{{j\neq
      i=1}}^{N_v}{\log{|{\bf r}_{i}{-}{\bf r}_{j}|}}.
\end{aligned}
\end{equation}
Here, $\rho_0(\lambda){=}\sqrt{2\lambda/\pi}$ is the density of the
condensate at the center of the trapping potential
and minimizing these energies will determines the positions
of vortices. In doing this, we will assume that the variations of
$\lambda$ are accompanied by an adiabatic change of the angular
frequency so that
${\Omega_{N_v}(\lambda)\le\Omega\ll\Omega_{N_v+1}(\lambda)}$, 
which
ensures that the wave-function minimizing the energy functional 
carries exactly $N_{v}$ vortices.
\begin{figure}[tb]
  \hskip0.1cm{{\bf (a)}\hskip4.0cm{\bf (b)}}\\
  \includegraphics[width=9cm]{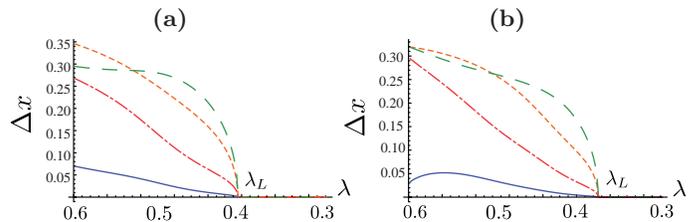}
  \caption{(Color online) Distance $\Delta{x}$ of the vortices from
    the soft trapping axis (in units of $\sqrt{2 N g \Omega/\hbar
      \omega_x}$) against the eccentricity of the trap,
    $\lambda$.  We show the cases of $N_v=7$ and $8$ (panel {\bf (a)}
    and {\bf (b)} respectively) and plot only the changes in positions
    of four vortices in the lattice (the association with the curves
    is irrelevant) , with the remaining showing analogous
      behavior.  At $\lambda=\lambda_{L}$ the vortices suddenly align
    along the y-axis ($\Delta x=0$). We have used a BEC of $10^6$
    $^{87}$Rb atoms with a scattering length
    $a=5.23\times{10}^{-9}$m in a trap of frequencies
    $\omega_z/2\pi=100$Hz and $\sqrt{\omega_x\omega_y}/2\pi=50$Hz
    (independent of $\lambda$).  }
\label{fig:delta}
\end{figure}
\begin{figure}[tb]
\includegraphics[width=\linewidth]{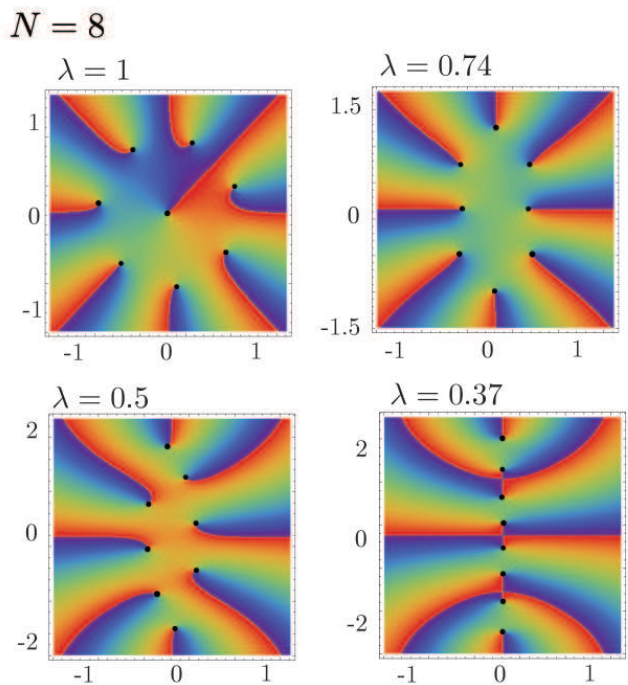}
\caption{(Color online) Phase distribution of the condensate carrying
  a vortex-lattice in the $x-y$ plane with $N_v=8$ for different
  values of $\lambda$.  The black dots mark the positions of the
  vortices (in units of $\sqrt{2 N g \Omega/\hbar \omega_x}$) The
  appearance of different structural vortex patterns is clearly
  visible.  }
\label{fig:N8}
\end{figure}
\begin{figure}[tb]
\includegraphics[width=\linewidth]{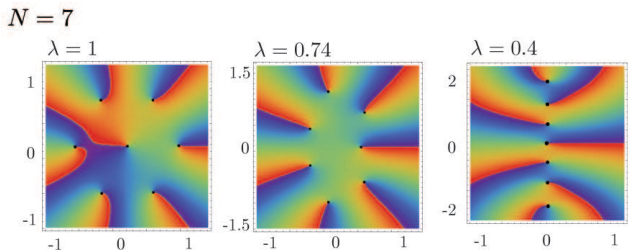}
\caption{(Color online) Phase distribution of the condensate carrying
  a vortex-lattice in the $x-y$ plane with $N_v=7$ for different
  values of $\lambda$. All other values are as in Fig.~\ref{fig:N8} }
\label{fig:N7}
\end{figure}
The absence of a $\Omega$-dependent term from the expression of $U_T$
can be understood by considering that, in the TF limit, the
centrifugal force 
is proportional to the restoring term for
$\Omega\approx\Omega_{N_v}(\lambda)$ (see Eq.~(3.4) in Ref. [14]), so
that they sum up to a quantity which is independent of $\Omega$.
In order to quantitatively asses the deviations of the vortex pattern
from the Abrikosov-like lattice~\cite{suzanne}, we first show how the
distances of the vortices from the tight trapping direction vary
against the eccentricity $\lambda$.  Two representative cases
($N_v=7,8$) of the general dynamics are shown in
  Fig.~\ref{fig:delta}: the pattern of vortices corresponding to
values of $\lambda$ larger than a {\it critical} threshold $\lambda_L$
(in general a function of $N_v$) abruptly collapses to an all-aligned
configuration.

\begin{figure}[tb]
\includegraphics[width=9 cm]{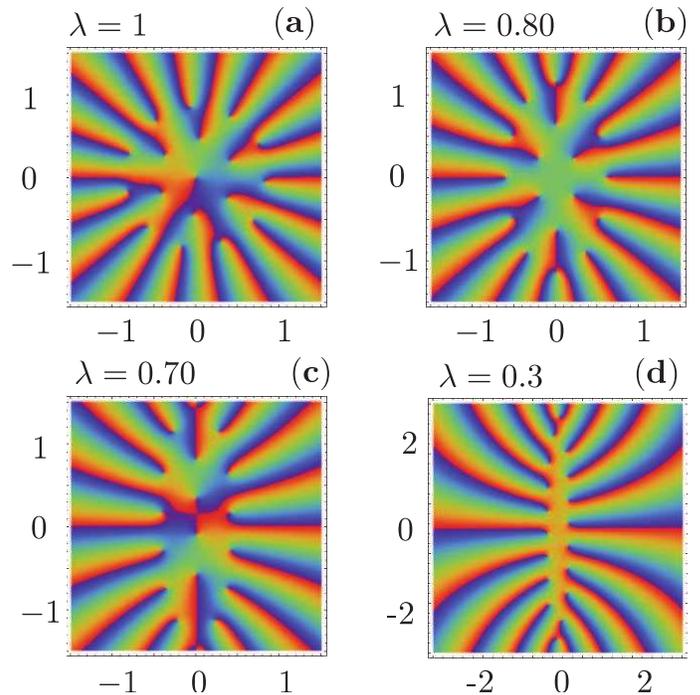}
\caption{(Color online) Phase distribution of the condensate carrying
  a vortex-lattice in the $x{-}y$ plane with $N_v{=}{18}$ for
  different values of $\lambda$.  The black dots mark the positions of
  the vortices (in units of $\sqrt{2 N g \Omega/\hbar \omega_x}$).  }
\label{fig:configs18}
\end{figure}

However, looking at the distance of the vortices from the soft axis
only gives limited information about the actual vortex pattern and we
show in Figs.~\ref{fig:N8} and \ref{fig:N7} the full position
distribution for different numbers of vortices and different values of
$1{\ge}\lambda{>}\lambda_L$. Two more structurally distinct
configurations become evident from this and let us first consider the
case of an even number of vortices (shown in Fig.~\ref{fig:N8}):
starting from an Abrikosov-like pattern at zero eccentricity
($\lambda=1$), the first structural change at ${\lambda=\lambda_C}$
witnesses the central vortex being displaced so as to join the ring
formed by the outer ones. A further reduction of $\lambda$ leads to a
second threshold value, $\lambda_Z$, at which the mirror symmetry is
broken and a zig-zag pattern is formed. The situation is different for
an odd number of vortices, where a parity effect leads to the
Abrikosov-to-ring and ring-to-zig-zag transitions becoming degenerate:
from full isotropy the lattice re-arranges directly into a zig-zag
pattern at $\lambda=\lambda_Z$, see Fig.~\ref{fig:N7}.  For even, as
well as odd number of vortices
a further reduction in $\lambda$ makes the vortices align along the
weak trapping direction, as already observed in Fig.~\ref{fig:delta}.
The situation is even richer for a larger (but finite) number of
vortices.  Let us consider, for instance, a system consisting of $18$
vortices (see Fig.~\ref{fig:configs18}).  As shown in panel (a), at
$\lambda=1$ they arrange in a pattern with a single vortex at the
centre of the trap and two concentric rings surrounding it.  By
decreasing $\lambda$ we first observe an Abrikosov-to-ring structural
change involving the inner ring (made out of six vortices) and the
central vortex, similar to the one described above (see panel (b)). By
further decreasing $\lambda$, the vortices in the newly formed inner
ring start joining the outer one (see panel (c)) before forming a
zig-zag pattern (panel (d)). Finally, the transition into a linear
structure occurs (not shown).

Let us briefly compare the vortex patterns we have just discussed with
the ones presented in Refs.~\cite{stripe,stripeII}. In these works the
authors address the case of an asymmetric trapping potential in the
fast rotation limit, $\Omega\rightarrow \min(\omega_x,\omega_y)$.  In
this case the condensate background cloud is stretched along the
direction of weak confinement and assumes a stripe-like shape. Because
of the symmetry of such a system vortices have to enter the cloud in
rows and the possible geometries are given by whether or not vortices
between different rows are aligned. In the case of fast rotating traps
the solution is determined by single particle states and if the
vorticity exceeds the number of atoms in the system, the existing
lattice {\it melts} and a highly correlated state emerges. In
contrast, we are dealing with a fixed number of vortices in the limit
where the interaction energy dominates the centrifugal one. For such
systems the healing length is much smaller than any other
characteristic length of the system, which, as we have shown, leads to
a number of possible patterns with well localised
singularities. Transitions between these patterns are then determined
by the interplay between the trapping potential and the interaction
energy between the vortices.


\begin{figure}[tb]
  \includegraphics[width=8 cm]{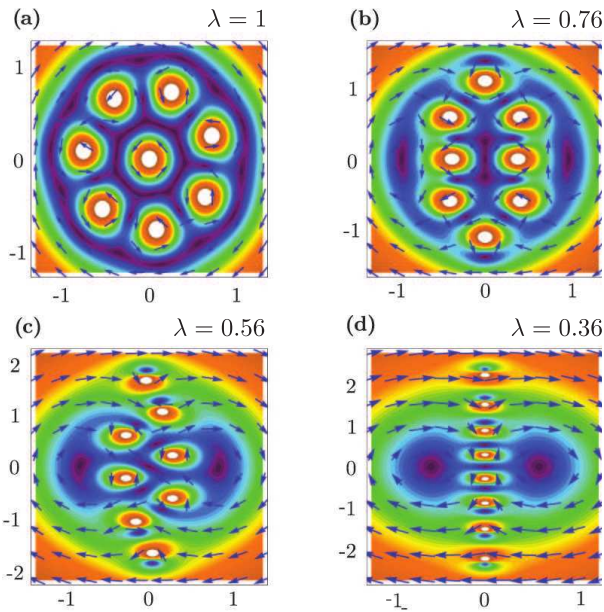}
  \caption{(Color online) Superfluid velocity field in the rotating
    frame for $N_v=8$ (other parameters as in
    Fig.~\ref{fig:delta}). From {\bf (a)} to {\bf (d)} the asymmetry
    parameter is given by $\lambda{=}1, 0.76, 0.56, 0.36$. Dark purple
    regions correspond to zero velocity and the velocities close to
    the vortex cores are not shown on the chosen
    color-map.}
\label{fig:velocity}
\end{figure}

\section {Superfluid hydrodynamics} 
In this Section we will explore the structural
transitions in detail by looking at the change in the
superfluid motion of the condensate. This is analogous to an argument
used by Fetter in Ref.~\cite{fetterell}, where superfluid motion in an elliptical and
{\it rectangular} cylinder was studied.
While both the energy and angular momentum
of the system were found~\cite{fetterell}, the existence of a threshold value
for the angular velocity above which the configuration with one vortex
is energetically favorable was shown. It is important to stress that in
  our case the vortex-lattice configuration found by minimizing
Eq.~\eqref{eq:inter} does not represent, in general, a {\it rigid
  pattern}, due to the perturbations introduced into the system by the
eccentricity. This can be seen by recasting the trapping
potential as 
\begin{equation}
V_{\lambda}({\bf x}){\equiv}V_{s}({\bf
  x}){+}V_{Q}(\lambda,y){=}\frac{1}{2}m\omega_x^2(x^2{+}y^2){+}\frac{1}{2}m\omega_x^2(\lambda^2{-}1)y^2
\end{equation}
and recognizing $V_{Q}(\lambda, y)$ as a term exciting quadrupole
modes.  Thus the background condensate and the vortex pattern are not
stationary.

The free energy of the rotating BEC is now given by
$F_{N_{v}}{=}E_{N_{v}} (\Omega,\lambda){+}U_T+U_I$ where $U_{T,I}$ are
defined by Eq.~\eqref{eq:inter} and $E_{N_{v}}
(\Omega,\lambda)$ is an energy term that does not depend on the vortex
configuration and whose detailed form is not essential for our
discussions.  By calling $\{{\bf r}_i^0\}$ (${i=1,..,N_v}$) the vortex
positions which minimize Eq.~(\ref{eq:inter}) for a set number of
vortices, we have the condition $\nabla_{j}
F_{N_{v}}\vert_{\{{\bf r}_i^0\}}{=}0$, where
$\nabla_j{\equiv}(\partial_{x_j},\partial_{y_j})$ and where we have
used the subscript $j$ to represent the coordinates of the
$j^\text{th}$ vortex. In the rotating frame, a vortex has a velocity
${\bm v}_j$ such that 
\begin{equation}
  \nabla_{j} F_{N_{v}}\vert_{{\bf r}_j}{\cdot}\bm{v}_{{\bf r}_j}{=}0,
\end{equation} 
which implies the absence of dissipation, as expected from particles
moving in a superfluid.  A solution to this equation is given by
${\bm{v}_{{\bf r}_j}{=}\alpha(\nabla_{j} ^{\perp}F_{N_{v}}\vert_{{\bf
      r}_j})}$ with $\nabla_{j}
^{\perp}{\equiv}(\partial_{y_{j}},-\partial_{x_{j}})$, where $\alpha$
is the amplitude of the velocity field. Its value
\begin{equation}
\alpha=a_{ho}\frac{\sqrt{\Omega \omega_x}}{\pi \rho_0(\lambda)}~~\left(\text{with}~
a_{ho}{=}\sqrt{\frac{\hbar}{m\omega_x}}\right)
\end{equation}
is found by comparing it with the
velocity field ${(\hbar/m)\nabla S{-}{\bf \Omega}{\times}{\bf r}_j}$
in the rotating frame. In this expression, $S{=}S_0{+}\sum_{i \neq
  j}^{N_v}\theta_{i}$ is the phase of the order parameter as seen by
the $j^{\text{th}}$ vortex, $\tan\theta_j{=}(y{-}y_j)/(x{-}x_j)$ specifies the polar angle of a
reference frame centered on the $j^\text{th}$ vortex
core~\cite{confenergy} and 
\begin{equation}
{S_0{=}-\frac{m\Omega(1{-}\lambda^2)}{\hbar(1{+}\lambda^2)}xy}
\end{equation}
 is the
vortex-free phase of the BEC at position $(x,y)$. 

In Fig.~\ref{fig:velocity} we show the magnitude of the velocity field
for $N_v=8$ in a frame which rotates rigidly with the trap. The value
of $\lambda$ decreases from panel {\bf (a)} to {\bf (d)} and the
arrows show the flow directions with the magnitude being encoded in
the color.  In the dark (dark purple) regions the velocity field
vanishes, {\it i.e.} the superfluid moves at the trap angular
velocity. For no eccentricity [panel {\bf (a)}] the vortex pattern
rotates rigidly with the trap potential since the velocity field at
the vortex positions (when the vortex itself is not present) vanishes
in the rotating frame.  It is worth noticing that outside the vortex
pattern particles flow with a different velocity.  This is at the
origin of the imperfect rigid-body rotation of finite-sized vortex
patterns in isotropic traps. By increasing the eccentricity [panel
{\bf (b)}-{\bf (d)}] the rigid body behavior is lost and the vortex
pattern is no longer a steady solution~\cite{suzanne}, since the
continuous rotation of the trap increases the angular momentum of the
system. However, the condition
$\Omega{\in}[{\Omega_{N_v},\Omega_{N_v+1}}[$ on the angular velocity
fixes the number of vortices in the condensate $N_v$. The only
possibility for the system to react is to move the vortex cores to
accommodate the angular momentum. In a real system, heating and
dissipation would eventually lead to the crystallization of the vortex
pattern or the transition to a turbulent regime \cite{cryst}.


\section {Vortex lattice modes} 
A quantitative confirmation of
the abrupt nature of the structural changes can be found by studying
the eigenmodes of the vortex pattern~\cite{campbell}. We take a set of
small displacements $\{\delta{\bf r}_i\}$ from the equilibrium
configuration ${\{{\bf r}_i^0\}}$ and write 
\begin{equation}
{\delta\bm{v}=(\delta
  v^{x}_{{\bf r}_1},\delta v^{y}_{{\bf r}_1},..,\delta v^{x}_{{\bf
      r}_{N_{v}}},\delta v^{y}_{{\bf r}_{N_{v}}})}, 
 \end{equation}
      so that
 {the vortex cores velocities in the rotating frame} become ${\delta\bm{v}\simeq{\bf A}\cdot{\bf
    \delta{\bf r}}}$. Here ${\bf A}$ is a $2 N_{v}{\times}2 N_{v}$
matrix whose $j^{\text{th}}$ row is found by expanding the velocity
field ${\bm{v}_{{\bf r}_j}{=}\alpha\nabla_{j}
  ^{\perp}F_{N_{v}}\vert_{{\bf r}_j}}$ around each ${\bf r}_i^0$. This gives 
  \begin{equation}
  {\bf
  A}_j{=}\alpha\sum_{i}\left[\partial_{x_{i}}(\nabla_{j}^{\perp}F_{N_{v}})\;
  \hat{x}_i\;+\;\partial_{y_{i}}(\nabla_j^{\perp}F_{N_{v}})\;\hat{y}_i
\right]_{\{{\bf r}_{i}^{0}\}},
\end{equation}
 where $\alpha$ is determined as
before. We now numerically diagonalize ${\bf A}$ for a set number 
of vortices.   
The eigenvalues $\alpha_l$ ($1{\le}l{\le}2 N_v$) of ${\bf A}$ represent the 
rate at which vortices start moving from ${\{{\bf r}_i^0\}}$ once they are 
displaced by the corresponding eigenvector $\delta {\bf r}^l$.
We note that the eigenmodes are related by 
\begin{equation}
\alpha_n (\lambda){+}\alpha_{2 N_v -n}(\lambda){=}C(\lambda)~~~~(0{<}n{\le}N_v)
\end{equation}
and the corresponding eigenvectors are mutually orthogonal. 
The constant $C(\lambda)$ depends on the system parameters but, remarkably, 
is independent of the pair of eigenvectors considered.
 \begin{figure}[tb]
  \includegraphics[width=\linewidth,]{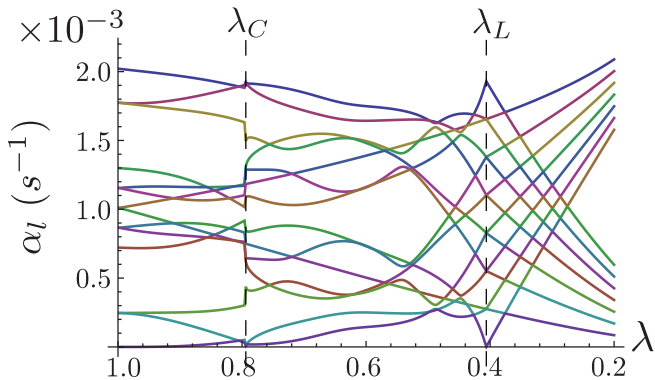}
  \caption{(Color online). Spectrum of a BEC with
    $N_v=7$ vortices against the  eccentricity $\lambda$. The
    points $\lambda_{C,L}$ where the vortex pattern
    undergoes a structural change are visible. }
\label{eigenmodes}
\end{figure}
A typical spectrum for $N_v{=}7$ is shown in Fig.~\ref{eigenmodes}. At
two specific values of $\lambda$ the eigenmodes show non-continuous
behavior, beside the appearance of a null eigenvalue.  These points
can be connected to the structural transition points: $\lambda_C$
signaling the Abrikosov-to-ring transition and $\lambda_L$ the
zig-zag-to-linear one. At any other value of $\lambda$ the eigenmodes
are positive confirming our previous point on the non-steady nature of
the vortex patterns in the rotating frame. However, the exact value of
$\lambda$ at which the lowest eigenvalue first deviates from zero is
found to grow with the number of vortices. The corresponding
eigenvector corresponds to displacements of the vortex positions along
the tangent to the vortex ring, {\it i.e.} a rotation of the vortex
pattern produces no effect.
In fact, it is not possible to clearly discriminate the eigenmodes of
a finite-size lattice with a small number of vortices from the phonon
modes of the background condensate: rotating an anisotropic trap
excites Bogoliubov modes in the BEC, which have a strong influence on
the vortex pattern~\cite{vortphn}.  The link between Bogoliubov modes
and changes in the properties of the {\it vortex matter} has already
been explored in relation to vortex-pattern formation and
instability~\cite{castin}.  Moreover, in Ref.~\cite{nonrotating}
stability of vortex clusters (comprising both vortices and
anti-vortices) in a non rotating anisotropic trap has been studied by
looking at the Bogoliubov modes.  The dynamics induced between the
background cloud and the vortex matter has there been shown to be not
separable.

\section{Conclusions} 
We have studied the structural transitions induced in a finite
vortex-lattice by an increasing degree of eccentricity of a rotating
BEC. An Abrikosov-like arrangement undergoes a sequence of
symmetry-breaking processes that push it towards a linear arrangement
of vortices. Such modifications, witnessed and understood in terms of
background superfluid motion, are well signaled by the eigenmodes of
the vortex-lattice. By addressing the case of a finite lattice, our
work complements and extends the existing literature on vortex
instabilities and arrangements in rotating BECs and provides
interesting insight into the many-body properties of a mesoscopic
quantum system.  Our analysis is not limited to BECs: vortex-like
excitations exist in superconducting films, Josephson-junction arrays
and dislocation pairs in the theory of 2D
melting~\cite{mullen}. Inter-vortex potentials depending
logarithmically on the distance between two vortices, similar to
Eq.~(\ref{eq:inter}), have been observed in thin superconducting
films~\cite{pearl}. Vortex lattices in thin films under magnetic
fields have been shown to take the form of discrete
rows~\cite{luzhbin}. Strong analogies between the dynamics of vortex
lattices and Josephson-junction arrays hold due to the charge-vortex
duality~\cite{bruder}, thus giving our results a generality and
interest that goes beyond the cases addressed here.

\acknowledgments

We thank G.~Morigi and W.~Bao for helpful discussions and invaluable
help. NLG thanks G.~Pucci for the Mini-amo project.  This work was
supported by SFI under grant numbers 05/IN/I852 and 05/IN/I852 NS,
IRCSET through the Embark Initiative (RS/2000/137) and EPSRC
(E/G004579/1).

\end{document}